\documentclass[twocolumn,showpacs,prl,aps,amssymb]{revtex4}

\usepackage{graphicx}% Include figure files
\usepackage{bm}% bold math

\begin{document}

\title{Spin-chirality duality in a spin ladder with four-spin
cyclic exchange}

\author{Toshiya Hikihara$^1$, Tsutomu Momoi$^2$, and Xiao Hu$^1$}
\affiliation{$^1$Computational Materials Science Center, National
Institute for Materials Science,
Tsukuba, Ibaraki 305-0047, Japan\\
$^2$Institute of Physics, University of Tsukuba, Tsukuba, Ibaraki
305-8571, Japan}
\date{\today}

\begin{abstract}
Effect of four-spin cyclic exchange on magnetism is studied in the
two-leg $S=1/2$ ladder. We develop an exact spin-chirality duality
transformation, under which the system is self-dual when the
four-spin exchange $J_4$ is half of the two-spin exchange. Using
the density-matrix renormalization-group method and the duality
relation, we find that the four-spin exchange makes the vector
chirality correlation dominant.
%we determine the ground-state phase diagram for the
%whole region of $0 \le J_4 \le \infty$. It is found that the
%four-spin exchange makes the vector chirality correlation dominant.
%A ^^ ^^ chirality-critical" region and
A ^^ ^^ chirality short-range resonating-valence-bond" phase is
identified for the first time at large $J_4$.
\end{abstract}

\maketitle

Recently, it has been realized that the two-leg $S=1/2$ ladder
compound
La$_x$Ca$_{14-x}$Cu$_{24}$O$_{41}$\cite{Brehmer,Matsuda,Schmidt,Nunner}
and two-dimensional (2D) antiferromagnet
La$_2$CuO$_4$\cite{HondaKW,Coldea} have a certain strength of
four-spin cyclic exchange interactions. Theoretically, four-spin
cyclic exchange emerges in the strong-coupling expansion of the
one-band Hubbard\cite{Takahashi} and $d$-$p$\cite{SchmidtT} models
as the leading correction to the nearest-neighbor two-spin
exchange. Cyclic exchanges were also found to be large in
magnetism of 2D quantum solids, e.g. solid ${}^3$He
films\cite{Roger} and Wigner crystals\cite{Okamoto}. The effect of
four-spin cyclic exchange on magnetism is, however, hardly
understood, since it has frustration by itself: the question of
{\it what type of magnetic ordering tends to be realized by the
four-spin exchange} is still unsettled. For example, in the
context of magnetism of solid {}$^3$He films, it was argued that
the four-spin exchange on the triangular lattice can induce scalar
chirality\cite{KuboM}, though finite-size system analysis could
not find evidence for such ordering, instead showing spin-liquid
ground states\cite{Misguich}.

To clarify magnetism induced by the four-spin cyclic exchange, we
consider the spin ladder. Spin ladder antiferromagnets have been
attracting extensive interest because they have a spin gap, a
short-range resonating-valence-bond (RVB) ground state, and
superconductivity upon doping\cite{DagottoR}.
%Since four-spin
%exchange strongly competes with two-spin exchange, it can induce
%novel exotic phases in both insulating magnets and doped systems.
%Regarding properties of doped systems and mechanism of high-T$_c$
%superconductivity, only two-spin exchange has been discussed in
%the $t$-$J$ model.
%On magnetism in the two-leg $S=1/2$ ladder it
On the two-leg $S=1/2$ ladder it was shown numerically that the
spin gap decreases rapidly with increasing the four-spin cyclic
exchange $J_4$\cite{Brehmer} and vanishes at a critical point,
$J_4/J \simeq 0.3$ where $J$ is the two-spin exchange
\cite{HondaH,HijiiN}. The nature of the new phase for large $J_4$
was not established.

%Spin ladder antiferromagnets have been attracting extensive
%interest because they are good theoretical models for various
%quasi-one-dimensional (1D) materials, and also because they have a
%spin gap, a short-range resonating-valence-bond (RVB) ground
%state, and superconductivity upon doping\cite{DagottoR}. Recently,
%it has been realized that a four-spin cyclic exchange $J_4$ is
%necessary for
%quantitative understanding of experimental results on spin
%gaps\cite{Brehmer,Matsuda}, Raman peaks\cite{Schmidt}, and optical
%conductivity\cite{Nunner}, observed in the cuprate ladder
%La$_x$Ca$_{14-x}$Cu$_{24}$O$_{41}$. The four-spin exchange was
%also found in the 2D antiferromagnet
%La$_2$CuO$_4$\cite{HondaKW,Coldea}, the parent material of
%high-T$_{\rm c}$ superconductors. Theoretically, the four-spin
%exchange emerges in a perturbation expansion of the one-band
%Hubbard\cite{Takahashi} and $d$-$p$\cite{SchmidtT} models as the
%leading correction to the nearest-neighbor two-spin exchange.
%Since the four-spin exchange competes with the two-spin exchange,
%which has been discussed intensively in the $t$-$J$ model for
%mechanism of high-T$_c$ superconductivity, it is important to
%clarify possible effects of the four-spin exchange on both
%magnetic and electronic properties.

In this letter, we show that the four-spin exchange in the two-leg
$S=1/2$ ladder has a tendency to induce a vector chirality
correlation. First, we describe an exact duality transformation
between the N\'eel-spin operator $\left({\bf s}_{1,l}-{\bf
s}_{2,l}\right)/2$ and the vector-chirality one ${\bf
s}_{1,l}\times{\bf s}_{2,l}$ on the rungs, where ${\bf s}_{\mu,l}$
is the spin operator at the site on the leg $\mu=1,2$ and the rung
$l$. The system is self-dual under this transformation at
$J_4/J=1/2$, where the N\'eel-spin and the chirality interchange
their roles: the former gives the most dominant correlation for
small $J_4$ while the latter does for large $J_4$. Using the
density-matrix renormalization-group (DMRG) method\cite{White1}
and the spin-chirality duality, we studied the ground-state phase
diagram of the ladder for the whole region of $0 \le J_4 \le
\infty$. We find the spin short-range RVB phase, an intermediate
phase with a very small spin gap, and a novel {\it chirality}
short-range RVB phase.
%Using the density-matrix renormalization-group (DMRG)
%method\cite{White1} and the spin-chirality duality, we have been able to
%determine the ground-state phase diagram of the ladder for the whole region
%of $0 \le J_4 \le \infty$. We find three different phases, i.e., the
%spin short-range RVB phase, a critical phase including a {\it
%chirality-critical} region where the vector-chirality correlation
%is dominant, and a novel {\it chirality} short-range RVB phase.
Our findings of exotic magnetic states with dominant
vector-chirality correlation at large $J_4$ suggest that the
four-spin exchange can induce exotic electronic states in doped
systems such as high-T$_c$ superconductors, whereas only two-spin
exchanges have been taken into account in $t$-$J$ models in
searching the mechanism.

Let us consider the Hamiltonian defined as
\begin{eqnarray}
{\cal H} &=& \sum_l \{ J_{\rm rung} {\bf s}_{1,l}\cdot{\bf
s}_{2,l}
  + J_{\rm leg} ({\bf s}_{1,l}\cdot{\bf s}_{1,l+1}\nonumber\\
& &+{\bf s}_{2,l}\cdot{\bf s}_{2,l+1})
  + J_4 K_l\},\label{eq:Hamiltonian}
\end{eqnarray}
where $J_{\rm rung}$ ($J_{\rm leg}$) denotes the two-spin exchange
constant on rungs (legs) and $K_l$ the four-spin cyclic exchange
on a plaquette $\{(1,l),(2,l),(2,l+1),(1,l+1)\}$,
\begin{eqnarray}
  K_l
  &=& {\bf s}_{1,l}\cdot{\bf s}_{2,l} + {\bf s}_{1,l+1}\cdot{\bf s}_{2,l+1}
+ {\bf s}_{1,l}\cdot{\bf s}_{1,l+1} \nonumber\\
&+& {\bf s}_{2,l}\cdot{\bf s}_{2,l+1} + {\bf s}_{1,l}\cdot{\bf
s}_{2,l+1} + {\bf s}_{2,l}\cdot{\bf s}_{1,l+1}
\nonumber\\
&+& 4({\bf s}_{1,l} \cdot {\bf s}_{2,l})
     ({\bf s}_{1,l+1} \cdot {\bf s}_{2,l+1}) \nonumber \\
&+& 4({\bf s}_{1,l} \cdot {\bf s}_{1,l+1})
     ({\bf s}_{2,l} \cdot {\bf s}_{2,l+1})  \nonumber \\
&-& 4({\bf s}_{1,l} \cdot {\bf s}_{2,l+1}) ({\bf s}_{2,l} \cdot
{\bf s}_{1,l+1}). \label{eq:P4}
\end{eqnarray}
All the coupling constants are assumed to be positive, $J_{\rm
rung}, J_{\rm leg}, J_4 > 0$. It is instructive to rewrite the
Hamiltonian (\ref{eq:Hamiltonian}) as
\begin{eqnarray} {\cal H} &=&
   (J_{\rm rung} + 2 J_4) \sum_l {\bf s}_{1,l} \cdot {\bf s}_{2,l}
\nonumber \\
&+& (J_{\rm leg} + J_4) \sum_l
        \sum_{\mu=1,2} {\bf s}_{\mu,l} \cdot {\bf s}_{\mu,l+1}
\nonumber \\
&+& J_4 \sum _l \left( {\bf s}_{1,l} \cdot {\bf s}_{2,l+1}
                              + {\bf s}_{1,l+1} \cdot {\bf s}_{2,l} \right)
\nonumber \\
&+& 4 J_4 \sum_l \left( {\bf s}_{1,l  } \cdot {\bf s}_{2,l  }
\right)
                  \left( {\bf s}_{1,l+1} \cdot {\bf s}_{2,l+1} \right)
\nonumber \\
&+& 4 J_4 \sum_l \left( {\bf s}_{1,l  } \times {\bf s}_{2,l  }
\right)
            \cdot \left( {\bf s}_{1,l+1} \times {\bf s}_{2,l+1} \right).
\label{eq:Ham}
\end{eqnarray}
An interesting contribution of the cyclic exchange appears in the
last term, which introduces a coupling between vector chiralities
on nearest-neighbor rungs. This term tends to induce non-zero
vector chiralities on every rung arranged in an antiparallel
pattern. Hence, one can naively expect that for large $J_4$ the
vector chirality becomes an important degree of freedom although
the frustration between the various terms in eq.\ (\ref{eq:Ham})
complicates the situation. We will show later that the
vector-chirality correlation indeed becomes dominant for large
$J_4$.

To elucidate the relation between the spin and chirality degrees
of freedom, we construct a duality transformation between them.
Let us begin with the commutation relations between the total
rung-spins ${\bf W}_l \equiv {\bf s}_{1,l} + {\bf s}_{2,l} $ and
the vector chirality ${\bf V}_l \equiv 2~{\bf s}_{1,l} \times {\bf
s}_{2,l}$ given by
\begin{eqnarray}
\left[ W^\alpha_l, W^\beta_{l'} \right] &=& i \epsilon^{\alpha
\beta \gamma} W^\gamma_l \delta_{l,l'},
\nonumber \\
\left[ V^\alpha_l, V^\beta_{l'} \right] &=& i \epsilon^{\alpha
\beta \gamma} W^\gamma_l \delta_{l,l'},
\nonumber \\
\left[ W^\alpha_l, V^\beta_{l'} \right] &=& i \epsilon^{\alpha
\beta \gamma} V^\gamma_l \delta_{l,l'}. \nonumber
\end{eqnarray}
We note that the commutation relations are identical to those
which hold between the angular momentum and the Runge-Lenz vector
of an electron system in a hydrogen atom. We can disentangle the
algebra by introducing new operators defined by
\begin{eqnarray}
{\bf S}_l &\equiv& \frac{1}{2} \left( {\bf W}_l - {\bf V}_l
\right)
              =   \frac{1}{2} \left( {\bf s}_{1,l} + {\bf s}_{2,l} \right)
                  - {\bf s}_{1,l} \times {\bf s}_{2,l},
\label{eq:difS} \\
{\bf T}_l &\equiv& \frac{1}{2} \left( {\bf W}_l + {\bf V}_l
\right)
              =   \frac{1}{2} \left( {\bf s}_{1,l} + {\bf s}_{2,l} \right)
                  + {\bf s}_{1,l} \times {\bf s}_{2,l}.
\label{eq:difT}
\end{eqnarray}
These operators obey the commutation relations,
\begin{eqnarray}
\left[ S^\alpha_l, S^\beta_{l'} \right]
&=& i \epsilon^{\alpha \beta \gamma} S^\gamma_l \delta_{l,l'}, \nonumber \\
\left[ T^\alpha_l, T^\beta_{l'} \right]
&=& i \epsilon^{\alpha \beta \gamma} T^\gamma_l \delta_{l,l'}, \nonumber \\
\left[ S^\alpha_l, T^\beta_{l'} \right] &=& 0, \nonumber
\end{eqnarray}
and satisfy ${\bf S}^2_l = {\bf T}^2_l = 3/4$. Thus, the new
operators ${\bf S}_l$ and ${\bf T}_l$ are $S = 1/2$ pseudo-spin
operators decoupling each other. It is interesting to note that
the original spins ${\bf s}_{1,l}$ and ${\bf s}_{2,l}$ may be
expressed in terms of ${\bf S}_l$ and ${\bf T}_l$ simply by
interchanging their roles in eqs.\ (\ref{eq:difS}) and
(\ref{eq:difT}), i.e., ${\bf s}_{1,l} =\frac{1}{2} \left( {\bf
S}_l + {\bf T}_l \right)
                  + {\bf S}_l \times {\bf T}_l$
and ${\bf s}_{2,l} = \frac{1}{2} \left( {\bf S}_l + {\bf T}_l
\right)
                  - {\bf S}_l \times {\bf T}_l$.
We hence call this a ^^ ^^ duality" transformation. The relations
between the original and new spin operators are summarized as
\begin{eqnarray}
{\bf s}_{1,l} + {\bf s}_{2,l} &=& {\bf S}_l + {\bf T}_l,
\nonumber \\
{\bf s}_{1,l} - {\bf s}_{2,l} &=& 2~ {\bf S}_l \times {\bf T}_l,
\nonumber \\
- 2~ {\bf s}_{1,l} \times {\bf s}_{2,l} &=& {\bf S}_l - {\bf T}_l,
\nonumber \\
{\bf s}_{1,l} \cdot {\bf s}_{2,l} &=& {\bf S}_l \cdot {\bf T}_l.
\nonumber
\end{eqnarray}
The transformation thereby exchanges the N\'eel-spin and the
vector chirality on the same rung.

In terms of the new spin operators, the Hamiltonian (\ref{eq:Ham})
is rewritten as
\begin{eqnarray}
\tilde{\cal H}
&=& (J_{\rm rung} + 2J_4) \sum_{l} {\bf S}_{l} \cdot {\bf T}_{l} \nonumber \\
&+& \left( \frac{J_{\rm leg}}{2} + 2 J_4 \right)
        \sum_{l} \left( {\bf S}_{l} \cdot {\bf S}_{l+1}
                      + {\bf T}_{l} \cdot {\bf T}_{l+1} \right) \nonumber \\
&+& \frac{J_{\rm leg}}{2} \sum_{l} \left( {\bf S}_{l} \cdot {\bf
T}_{l+1}
                               + {\bf T}_{l} \cdot {\bf S}_{l+1} \right)
                               \nonumber \\
&+& 4 J_4 \sum_{l} \left( {\bf S}_{l  } \cdot {\bf T}_{l  }
\right)
                   \left( {\bf S}_{l+1} \cdot {\bf T}_{l+1} \right)
                   \nonumber \\
&+& 2 J_{\rm leg} \sum_{l} \left( {\bf S}_{l  } \times {\bf T}_{l
} \right)
                 \cdot \left( {\bf S}_{l+1} \times {\bf T}_{l+1} \right).
\label{eq:tHam}
\end{eqnarray}
Thus, the duality transformation leaves the form of the
Hamiltonian unchanged and only affects the coefficients of the
second, third, and fifth terms. An interesting observation here is
that in the case of $J_4 = J_{\rm leg}/2$ the original Hamiltonian
${\cal H}$ and its dual $\tilde{\cal H}$ are equivalent including
the coefficients. Hence the N\'eel-spin $\left( {\bf s}_{1,l} -
{\bf s}_{2,l} \right)/2$ and the vector chirality ${\bf s}_{1,l}
\times {\bf s}_{2,l}$ show identical behavior on 
this ^^ ^^ self-dual" line.

To clarify the consequence of the spin-chirality duality around
$J_4=J_{\rm leg}/2$ and the nature of ground states, we study the
low-energy properties of the Hamiltonian (\ref{eq:Ham})
numerically. For simplicity, we focus on the case of $J_{\rm rung}
= J_{\rm leg} = J$ and investigate the ground-state phase diagram
on the $J_4/J$ line hereafter. Using the DMRG method\cite{White1},
we have calculated the energy gap of spin excitations
\begin{equation}
\Delta_{0M} (L) = E_0(L;M) - E_0(L;0), ~~~~(M =
1,2),\label{eq:gap}
\end{equation}
where $E_0(L;M)$ is the lowest energy in the subspace of $s^z_{\rm
total} = \sum_l (s^z_{1,l}+s^z_{2,l}) = M$ in a finite ladder of
$L$ rungs. For the best performance of the DMRG method, an open
boundary condition was imposed. We have also calculated the
ground-state spin correlation functions
\begin{eqnarray}
C^s_0 (r) &=& \frac{1}{4} \langle \left( s^z_{1,l } + s^z_{2,l }
\right)
         \left( s^z_{1,l'} + s^z_{2,l'} \right) \rangle,
\label{eq:cor0} \\
C^s_\pi (r) &=& \frac{1}{4} \langle \left( s^z_{1,l } - s^z_{2,l }
\right)
         \left( s^z_{1,l'} - s^z_{2,l'} \right) \rangle,
\label{eq:corpi}
\end{eqnarray}
and the vector-chirality correlation function
\begin{equation}
C^\kappa (r) = \langle \left( {\bf s}_{1,l } \times {\bf s}_{2,l }
\right)^z
         \left( {\bf s}_{1,l'} \times {\bf s}_{2,l'} \right)^z \rangle,
\label{eq:vch}
\end{equation}
with $l = l_0-r/2$ and $l' = l_0+r/2$\cite{Stot}. The index $l_0$
represents the center position of the open ladder, i.e., $l_0 =
L/2$ for even $r$ and $l_0 = (L+1)/2$ for odd $r$. We have
employed the finite-system method with improved
algorithm\cite{White1} and kept up to $m = 500$ states per block.
The numerical errors due to the truncation are estimated from the
difference between the data of different $m$'s. The system size is
up to $2 \times 100$ sites.

\begin{figure}
   \includegraphics[width=86mm]{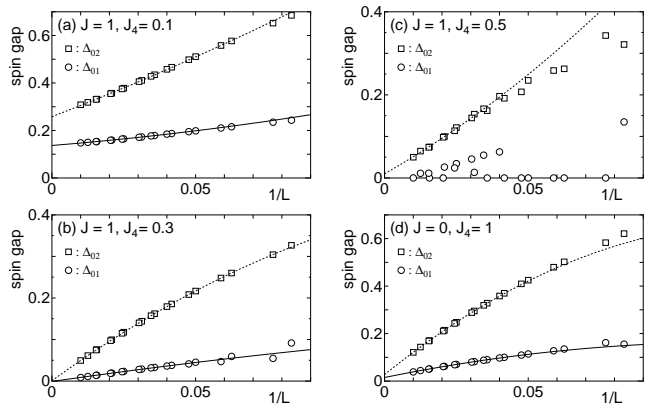}
\caption{System-size dependence of the spin gaps for (a) $(J,J_4)
= (1, 0.1)$; (b) $(1, 0.3)$; (c) $(1, 0.5)$; and (d) $(0,1)$. The
gaps $\Delta_{01}(L)$ and $\Delta_{02}(L)$ are plotted by circles
and squares, respectively. The numerical errors of the DMRG
calculation are smaller than the symbols.} \label{fig:gap-L}
\end{figure}
\begin{figure}
   \includegraphics[width=80mm]{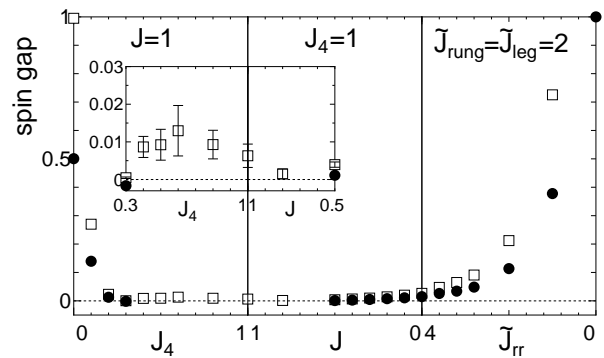}
\caption{Extrapolated spin gaps in the limit $L \to \infty$ as
functions of $J_4$ (left), $J$ (center), and $\tilde{J}_{\rm rr}$
(right); see text. The circles and squares represent
$\Delta_{01}(\infty)$ and $\Delta_{02}(\infty)$, respectively.
Inset: Enlarged figure for $0.3 \le J_4/J \le 2$. The error bars
represent those from the extrapolation procedure.} \label{fig:gap}
\end{figure}

First, we discuss the parameter region $J_4/J < 1$. It is known
that for $J_4 = 0$ the system belongs to a spin-liquid phase, in
which the ground state is well described by the RVB state. The
spin gap is open in this phase. We show in Fig. \ref{fig:gap-L}
our numerical results for the spin gaps $\Delta_{0M} (L)$. The
data are extrapolated by fitting them to a polynomial form,
$\Delta_{0M}(L) = \Delta_{0M}(\infty) + a/L + b/L^2$. For $J_4/J
\lesssim 0.3$, the spin gaps decrease smoothly for both $M$'s as
$L$ increases, and consequently, the extrapolation works pretty
well [see Fig.\ \ref{fig:gap-L}(a) and (b)]. The extrapolated
values $\Delta_{0M}(\infty)$ are shown in Fig.\ \ref{fig:gap} with
$J = 1$ in the region $J_4/J<1$. The spin gaps decrease smoothly
as $J_4$ increases from $0$ and vanish around $J_4/J \simeq 0.3$,
suggesting a phase transition accompanied by vanishing of the spin
gap at $J_4/J = (J_4/J)_{\rm c1} \simeq 0.3$. Unfortunately,
accurate estimation of the critical value $(J_4/J)_{\rm c1}$ is
quite difficult due to the very slow vanishing of the spin gaps
around the transition point. When $J_4/J$ is larger than $0.3$,
$\Delta_{01}(L)$ shows bumpy behavior as seen in Fig.\
\ref{fig:gap-L}(c). This may be attributed to effects of open
boundaries. The value of $\Delta_{01}(L)$ becomes exactly $0$
within numerical accuracy for several $L$, which suggests a
spin-triplet ground state. On the other hand, the spin gap
$\Delta_{02}(L)$ exhibits a rather smooth $L$-dependence even for
$J_4/J \ge 0.3$. The extrapolated gaps $\Delta_{02}(\infty)$ are
very small, less than $0.02$, but seem to be finite for this
region. Very recently, L\"auchli {\it et al.} studied
independently the same model but with a different boundary
condition and showed that the system for this parameter region
belongs to a phase with a very small gap exhibiting the
translational symmetry breaking\cite{Lauchli}. The work of Ref.\
\onlinecite{Muller} also pointed toward this result. Our results
are thus consistent with theirs although the number and  type of
excitations in the finite systems might differ from each other
because of the different boundary conditions. Further studies,
especially by analytic methods, are desirable for clarifying the
nature of excitations and why the spin gap is so small in the
entire region of the phase.
%When $J_4/J$ is larger than $0.3$, $\Delta_{01}(L)$ shows bumpy behavior as
%seen in Fig.\ \ref{fig:gap-L}(c).
%This may be attributed to an incommensurate character of the
%system and the finiteness of system size.
%Note that the value of $\Delta_{01}(L)$ becomes exactly $0$ within numerical
%accuracy for several $L$, which suggests a spin-triplet ground state.
%On the other hand, the spin gap $\Delta_{02}(L)$
%exhibits a rather smooth $L$-dependence even for $J_4/J \ge 0.3$
%and is extrapolated to zero in the limit $L \to \infty$.
%This result indicates gapless magnetic excitations above the ground state.
% From this observation, we conclude that there is a phase transition from
%a gapped to a critical phase at $J_4/J = (J_4/J)_{\rm c1} \simeq 0.3$.
%We expect that the appearance of the spin-triplet ground state is a
%finite-size effect and the system in the thermodynamic limit possesses
%a unique ground state with gapless excitations.
%Unfortunately, accurate estimation of the critical value
%$(J_4/J)_{\rm c1}$ is quite difficult due to the very slow vanishing
%of the spin gaps around the transition point.
%Our results on the spin gap $\Delta_{01}$ are in
%agreement with Ref.\ \onlinecite{HondaH}.

\begin{figure}
   \includegraphics[width=86mm]{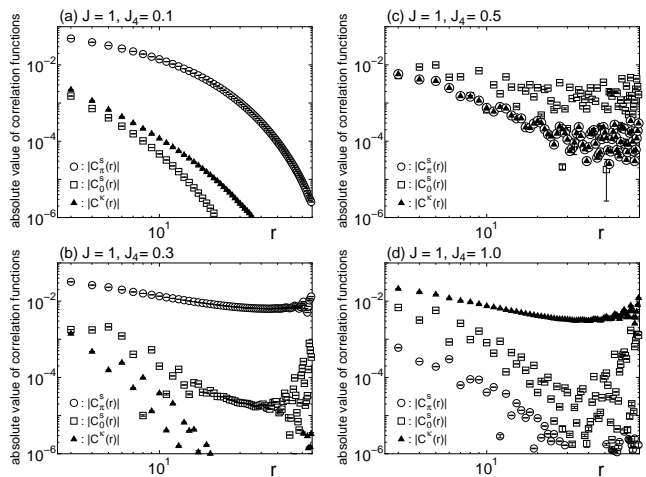}
\caption{Correlation functions $|C^s_\pi(r)|$ (circles),
$|C^s_0(r)|$ (squares), and $|C^\kappa(r)|$ (triangles) as
functions of distance for $J = 1$ and (a) $J_4 = 0.1$; (b) $J_4 =
0.3$; (c) $J_4 = 0.5$; and (d) $J_4 = 1$. The system size is $L =
80$. The error bars represent the numerical errors of the DMRG
calculation. } \label{fig:cor}
\end{figure}

Figure \ref{fig:cor} shows the spin correlations $C^s_0(r)$ and
$C^s_\pi(r)$, and the vector-chirality correlation $C^\kappa(r)$
for several typical sets of parameters. For $J_4/J < (J_4/J)_{\rm
c1} \simeq 0.3$, all the correlations decay exponentially, which
is consistent with a finite spin gap. The N\'eel-spin correlation
$C^s_\pi(r)$ is the strongest among the calculated correlation
functions [Fig.\ \ref{fig:cor}(a)], as in the usual
antiferromagnetic (AF) ladder.
%For $J_4/J > (J_4/J)_{\rm c1}$, on the other hand,
%the correlation functions
%decay algebraically\cite{open}, again being
%consistent with the absence of the spin gap.
%We note that the
%long-range order (LRO) of the spin and vector-chirality
%correlations is forbidden in 1D isotropic systems by a rigorous
%argument\cite{Momoi}, but their quasi LROs are allowed.
For $J_4/J > (J_4/J)_{\rm c1}$, on the other hand, the correlation
functions decay very slowly reflecting the small spin
gap\cite{open}. We find that the N\'eel-spin correlation
$C^s_\pi(r)$, which is dominant for small $J_4$, keeps reducing as
$J_4$ increases while the vector-chirality correlation $C^\kappa
(r)$ keeps growing and becomes dominant for large $J_4$. They
interchange with each other at the self-dual point $J_4/J=0.5$; we
can see their perfect coincidence in Fig.\ \ref{fig:cor}(c). These
results strongly suggest that the system exhibits symmetric
behaviors with respect to the self-dual point $J_4/J = 0.5$ with
exchanging the roles of the N\'eel-spin and the vector-chirality
correlations. We also note that the dimer operator $\vec{s}_{1,l}
\cdot \vec{s}_{1,l+1} - \vec{s}_{2,l} \cdot \vec{s}_{2,l+1}$ and
the scalar-chirality operator $(\vec{s}_{1,l} + \vec{s}_{2,l})
\cdot (\vec{s}_{1,l+1} \times \vec{s}_{2,l+1}) + (\vec{s}_{1,l}
\times \vec{s}_{2,l}) \cdot (\vec{s}_{1,l+1} + \vec{s}_{2,l+1})$
are related to each other by the duality transformation, and
consequently, their correlation functions must interchange exactly
at $J_4/J = 0.5$, which is consistent with numerical results in
Ref.\ \onlinecite{Lauchli}.
%In a narrow intermediate region around the self-dual point, we also
%find that the total-rung-spin correlation $C^s_0 (r)$ becomes the
%strongest. The critical region is thus divided into three regions,
%i.e., the N\'eel-spin-critical [$(J_4/J)_{\rm c1} < J_4/J \lesssim
%0.45$], total-rung-spin-critical ($0.45 \lesssim J_4/J \lesssim
%0.6$), and {\it chirality-critical} ($J_4/J \gtrsim 0.6$) regions
%where the correlations $C^s_\pi (r)$, $C^s_0 (r)$, and $C^\kappa
%(r)$ are dominant, respectively.

Next, we consider the parameter region $J_4/J > 1$. Hereafter, we
set $J_4 = 1$. It can be seen in Fig.\ \ref{fig:gap} that the spin
gaps open for $J_4/J> (J_4/J)_{\rm c2}\simeq 1/0.4 = 2.5$. Again,
accurate estimation of $(J_4/J)_{\rm c2}$ is quite difficult due
to the very slow opening of the gaps. We also note that for large
$J_4/J$ the spin gap $\Delta_{01}$ exhibits a smooth
$L$-dependence and does not become $0$ for finite $L$ [see Fig.\
\ref{fig:gap-L}(d)], suggesting the absence of the triplet ground
state in the finite systems. To elucidate the nature of the system
in this large $J_4/J$ region, we consider the case $J_4/J=\infty$
($J_4 = 1$ and $J = 0$) using the spin-chirality duality
transformation. In this case, the transformed Hamiltonian
(\ref{eq:tHam}) is expressed as
\begin{eqnarray}
\tilde{\cal H}_{J_4} &=& \tilde{J}_{\rm rung} \sum_{l} {\bf S}_{l}
\cdot {\bf T}_{l}
  +  \tilde{J}_{\rm leg}
        \sum_{l} \left( {\bf S}_{l} \cdot {\bf S}_{l+1}
                      + {\bf T}_{l} \cdot {\bf T}_{l+1} \right) \nonumber \\
&+& \tilde{J}_{\rm rr} \sum_{l}
                         \left( {\bf S}_{l  } \cdot {\bf T}_{l  } \right)
                         \left( {\bf S}_{l+1} \cdot {\bf T}_{l+1} \right)
\label{eq:HamJ4}
\end{eqnarray}
with $\tilde{J}_{\rm rung} = \tilde{J}_{\rm leg} = 2$ and
$\tilde{J}_{\rm rr} = 4$. Notice here that, if one sets
$\tilde{J}_{\rm rr} = 0$ in eq.\ (\ref{eq:HamJ4}), the system is
equivalent to the usual two-leg AF ladder, which has the
short-range RVB ground state consisting of the spins ${\bf S}_l$
and ${\bf T}_l$. In Fig. \ref{fig:gap}, we show the
$\tilde{J}_{\rm rr}$-dependence of the extrapolated spin gaps
$\Delta_{0M}(\infty)$ for $0 \le \tilde{J}_{\rm rr} \le 4$. It is
clear that the spin gaps remain finite for the entire region of $0
\le \tilde{J}_{\rm rr} \le 4$ and are smoothly connected to the
spin gaps at $\tilde{J}_{\rm rr} = 4$; there is no phase
transition between $\tilde{J}_{\rm rr} = 0$ and $\tilde{J}_{\rm
rr} = 4$. We thus conclude that the Hamiltonian (\ref{eq:HamJ4})
with $\tilde{J}_{\rm rr} = 4$, and accordingly, the original
ladder (\ref{eq:Ham}) in the limit $J_4/J=\infty$ belong to the
same RVB phase as the AF ladder of the spins ${\bf S}_l$ and ${\bf
T}_l$ with $\tilde{J}_{\rm rr} = 0$. Small size of the spin gap at
$J_4/J=\infty$ can be understood from the fact that the system is
close to the quantum critical point between the short-range RVB
and intermediate phases. Note that the dominant correlation
function in this RVB phase is that of the N\'eel-spin $\left( {\bf
S}_l - {\bf T}_l \right)/2$ and hence, in terms of the original
spins, the correlation of the vector chirality ${\bf s}_{1,l}
\times {\bf s}_{2,l}$ is the strongest. We therefore term this
novel phase the {\it chirality} short-range RVB phase.

%To summarize, we have developed the exact duality transformation
%between the N\'eel-type spin and the vector chirality,
%under which the system is self-dual at $J_4/J = 0.5$.
%The four-spin cyclic exchange
%makes the vector chirality correlation dominant.
%We have found that he chirality RVB phase appear for large $J_4$.
To summarize, using the spin-chirality duality transformation,
which is developed in this letter, as well as the DMRG method, we
have found that the four-spin cyclic exchange makes the vector
chirality correlation dominant. The chirality RVB phase appears
for large $J_4$. It has been found that the system exhibits
symmetric behavior with respect to the self-dual point $J_4/J =
1/2$ by interchanging the N\'eel spin and the vector chirality. We
remark that the duality transformation is applicable to any spin
Hamiltonian, since it is based only on the spin commutation
relation. This transformation should be useful in studying various
topics. One example is the spin-orbital model around the SU(4)
symmetric point\cite{Yamashita}. We have found in the two-leg
ladder with extended four-spin exchange that the self-dual line
connects with the SU(4)-symmetric point\cite{HikiharaMH2}.
%We expect that the critical phase in the present ladder is
%controlled by the SU(4) fixed point\cite{Azaria}.
Another example is a magnetization plateau induced by the
four-spin exchange\cite{Sakai}. Since the total spin $\sum_{\mu,l}
{\bf s}_{\mu,l}$ is invariant under the dual transformation, the
duality relation holds even in a magnetic field. Effect of
four-spin exchange on hole-doped systems is also to be considered.
It would be interesting to investigate relation to possible hidden
orders proposed for high-T$_{\rm c}$ superconductors, e.g. the
staggered currents.

We would like to thank K.\ Kubo, N.\ Taniguchi, A.\ Tanaka, K.\
Nomura and M.\ Nakamura for stimulating discussions. We also thank
A.\ L\"auchli for useful comments. T.M. was supported by Monkashou
(MEXT) in Japan through Grant Nos.\ 13740201 and 1540362.

%\vspace*{-3mm}


\begin{thebibliography}{99}
\bibitem{Stot}
To be precise, the correlation functions are calculated in the
lowest-energy state in the subspace $s^z_{\rm total} =0$, which
must be the ground state or, at least, one of the ground states
since the Hamiltonian is $SU(2)$ symmetric.
\bibitem{open}
We note that the enhancement of the correlations seen for large $r
\sim L$ is due to the presence of the open boundaries and is not
essential.
\end{thebibliography}
\end{document}